\documentstyle[twocolumn,aps,epsf]{revtex}
\begin{document}
%\draft
%\input{epsf}
\title{Mass dependence of light nucleus production in ultrarelativistic 
heavy ion collisions}
%\maketitle
\author{
T.A. Armstrong                \unskip,$^{(8,\ast)}$
K.N. Barish                   \unskip,$^{(3)}$
S. Batsouli                  \unskip,$^{(13)}$
S.J. Bennett                  \unskip,$^{(12)}$
A. Chikanian                  \unskip,$^{(13)}$
S.D. Coe                      \unskip,$^{(13,\dag)}$\\
T.M. Cormier                  \unskip,$^{(12)}$
R. Davies                     \unskip,$^{(9,\ddag)}$
C.B. Dover                    \unskip,$^{(1,\S)}$
P. Fachini                    \unskip,$^{(12)}$
B. Fadem                      \unskip,$^{(5)}$
L.E. Finch                    \unskip,$^{(13)}$
N.K. George                   \unskip,$^{(13)}$\\
S.V. Greene                   \unskip,$^{(11)}$
P. Haridas                    \unskip,$^{(7)}$
J.C. Hill                     \unskip,$^{(5)}$
A.S. Hirsch                   \unskip,$^{(9)}$
R. Hoversten                  \unskip,$^{(5)}$
H.Z. Huang                    \unskip,$^{(2)}$
H.   Jaradat                  \unskip,$^{(12)}$\\
B.S. Kumar                    \unskip,$^{(13,\|)}$
T. Lainis	              \unskip,$^{(10)}$
J.G. Lajoie                   \unskip,$^{(5)}$
R.A. Lewis                    \unskip,$^{(8)}$
Q. Li                         \unskip,$^{(12)}$
B. Libby                      \unskip,$^{(5,\P)}$
R.D. Majka                    \unskip,$^{(13)}$
T.E. Miller                   \unskip,$^{(11)}$\\
M.G. Munhoz                   \unskip,$^{(12)}$
J.L. Nagle                    \unskip,$^{(4)}$
I.A. Pless                    \unskip,$^{(7)}$
J.K. Pope                     \unskip,$^{(13,\natural)}$
N.T. Porile                   \unskip,$^{(9)}$
C.A. Pruneau                  \unskip,$^{(12)}$
M.S.Z. Rabin                  \unskip,$^{(6)}$\\
J.D. Reid                     \unskip,$^{(11)}$
A. Rimai                      \unskip,$^{(9,\sharp)}$
A. Rose                       \unskip,$^{(11)}$
F.S. Rotondo                  \unskip,$^{(13,\ell)}$
J. Sandweiss                  \unskip,$^{(13)}$
R.P. Scharenberg              \unskip,$^{(9)}$
A.J. Slaughter                \unskip,$^{(13)}$\\
G.A. Smith                    \unskip,$^{(8)}$
M.L. Tincknell                \unskip,$^{(9,\dag\dag)}$
W.S. Toothacker               \unskip,$^{(8)}$
G. Van Buren                  \unskip,$^{(7,\P\P)}$
F.K. Wohn                     \unskip,$^{(5)}$
Z. Xu                         \unskip,$^{(13)}$
}
\address{\centerline{(The E864 Collaboration)}}
\address{  $^{(1)}$ Brookhaven National Laboratory, Upton, 
New York 11973 \break
  $^{(2)}$ University of California at Los Angeles, Los Angeles, 
California 90095 \break  
  $^{(3)}$ University of California at Riverside, Riverside, 
California 92521 \break
  $^{(4)}$ Columbia University, Nevis Laboratory, Irvington, NY 10533\break
  $^{(5)}$ Iowa State University, Ames, Iowa 50011 \break 
  $^{(6)}$ University of Massachusetts, Amherst, Massachusetts 01003 \break 
  $^{(7)}$ Massachusetts Institute of Technology, Cambridge, 
Massachusetts 02139 \break 
  $^{(8)}$ Pennsylvania State University, University Park, 
Pennsylvania 16802 \break 
  $^{(9)}$ Purdue University, West Lafayette, Indiana 47907 \break 
  $^{(10)}$ United States Military Academy, West Point \break
  $^{(11)}$ Vanderbilt University, Nashville, Tennessee 37235 \break 
  $^{(12)}$ Wayne State University, Detroit, Michigan 48201 \break 
  $^{(13)}$ Yale University, New Haven, Connecticut 06520 \break
}
\date{\today}
\maketitle
\begin{abstract}
  Light nuclei can be produced in the central reaction zone via coalescence 
in relativistic heavy ion collisions. E864 at BNL has measured the 
production of ten light nuclei with nuclear number of $A=1$ to $A=7$ at 
rapidity $y\simeq1.9$ and $p_{T}/A\leq300MeV/c$. 
Data were taken with a Au beam of momentum of $11.5$ A $GeV/c$ on a Pb or Pt 
target with different experimental settings. 
The invariant yields show a striking exponential dependence on nuclear number 
with a penalty factor of about 50 per additional nucleon. Detailed analysis 
reveals that the production may depend on the spin factor of the nucleus and 
the nuclear binding energy as well. 
\end{abstract}
%\pacs{}
%\maketitle
  Relativistic heavy ion collisions may create high energy density and 
high baryon density in the reaction zone. Therefore, they are considered 
to be the best laboratory environment to create novel objects or states such 
as the Quark-Gluon Plasma\cite{jwh}. At the AGS energies ($\sqrt{s}\simeq4.8$
A$GeV$), the baryon stopping is high, which means that 
the baryon density near center-of-mass rapidity is quite large compared to 
the much higher beam energy at the SPS ($\sqrt{s}\simeq17$A$GeV$) or at 
the RHIC ($\sqrt{s}\simeq200$A$GeV$)\cite{stopping}. 
Light nuclei can be produced 
by the coalescence of created or stopped nucleons\cite{nagle}. 
%This recombination process is called coalescence. 
Since the probability of coalescence of a particular nuclear system 
(d, $^{3}He$, etc.) depends on the properties of the hadronic system formed 
as a result of the collision, the study of the coalescence process is useful 
in elucidating those properties. For example, in a coalescence model, the 
coalescence probability depends on the temperature, baryon chemical potential 
(essentially the baryon density), and the size of the system, as well as the 
statistical weight of the coalesced nucleus\cite{heinz}. 
A thermal model\cite{braun} does not use details of cluster creation. 
however, the dependence of the production on the system configuration is 
very similar to the coalescence model. The data presented in this paper 
shows evidence that the probability 
may also depend on the binding energy of the coalesced nucleus. 
Systematic study of the production of light nuclei is limited 
by their low production rates in relativistic heavy ion collisions. 
E864 is the only experiment which is able to measure the production of charged 
nuclei with $A>4$ produced in the central reaction zone. However, the 
nature of the coalescence process tells us that the higher the baryon number, 
the more sensitive the production rate is to the system's configuration. 

  In this letter, the production of protons, neutrons, deuterons, $^{3}He$, 
tritons, $^{4}He$, $^{6}He$, $^{6}Li$, $^{7}Li$ and $^{7}Be$ around rapidity 
$y\simeq1.9$ and transverse momentum of $p_{T}/A\leq300MeV/c$ in 10\% most 
central Au+Pt(Pb) collisions measured by E864\cite{nim} at BNL is  
presented. These measurements have significant impact on the strange 
quark matter\cite{jaffe} search by several experiments in relativistic 
heavy ion collisions. They also provide information about the thermal 
equilibrium of the system and the detailed process of coalescence. 
%\section{E864 Apparatus}

  E864\cite{nim} at Brookhaven National Laboratory is an open geometry, 
high data rate apparatus designed to search for novel objects such as 
strange quark matter. It is a fixed-target experiment at the AGS with 
incident Au beam momentum of $11.5GeV/c$ per nucleon. 
% More detailed descriptions of the apparatus can 
%be found in other publications\cite{calo,nim,haridas,let}.
We can have two independent mass 
measurements with the E864 apparatus:% as shown in Figure\ref{fig:e864}: 
the tracking system and the hadronic calorimeter. 
%They identify particles and reject background 
%powerfully with the confirmation of each other. 
The requirement that the two measurements agree provides excellent particle 
identification and background rejection. 
The tracking system has 
two dipole analyzing magnets (M1 and M2) followed by 
three hodoscope planes (H1, H2 and H3). There are two straw stations 
(S2 and S3), each with three close-packed double planes. 
Each scintillating hodoscope plane has 206 vertical slats, and there are 
960 and 1920 straw tubes in each S2 and S3 straw plane respectively. 
The tracking system measures 
momentum, charge and velocity (\(\beta\)) of charged particles with a mass 
resolution of few percents in the region of interest. Charge misidentification 
is less than 1 in $10^{8}$ due to three redundant measurements of energy 
loss in the hodoscopes.
%charge measurements from the scintillating hodoscopes' $dE/dX$ measurements.
There is an additional mass measurement from the hadronic 
calorimeter\cite{calo} 
which has good energy (\(\Delta E/{\sqrt{E}}={0.344/{\sqrt{E}}}+0.035\))  and 
time (\(\sigma_{t} \simeq 400ps\)) resolutions. It is made of 
754 towers of scintillating fiber-embedded lead. The containment of a 
hadronic shower is about 50\% in the peak tower and about 90\% in the 
$3\times3$ array around the peak tower. This allows sufficient quality 
requirements of the shower cluster and provides for good neutron measurements. 
A vacuum tank along the beam line reduces the background from beam particles 
interacting with air. The total length of the apparatus is about 28 meters.

The trigger consists of good beam definition, a multiplicity 
requirement\cite{haridas} and an optional level II high mass 
trigger -- the Late-Energy Trigger (LET)\cite{let}. The LET utilizes the 
energy and time-of-flight measurements from the calorimeter.
%to setup the level II trigger. 
Only the 10\% most 
central collisions were collected for the data presented here. The data 
are from four different experimental settings optimized for the different 
physics topics. Proton, deuteron, 
$^{3}He$ and triton measurements are from the 1995 run with the "+.45T" field 
setting on M1 and M2 magnets\cite{nigel} and with a Pb target of 5\% 
interaction length for Au nuclei. Neutron data are from the 1995 run with the 
highest field setting of "+1.5T" (5\% Pb target)\cite{evan} 
while the rest of the data are from the 1996 run at the "+1.5T" field setting 
with the high mass level II trigger and Pt target of 60\% interaction length 
(physical thickness of 1.5cm) mainly for the strange quark matter 
search\cite{xzb} with the exception that the $^{7}Be$ and $^{7}Li$ 
measurements were from the combination of "-.75T" and "+1.5T" 
field settings taken in the 1996 run. Due to the open geometry, there is 
sufficient overlap from different settings for consistency 
checks\cite{prc,pope,nigel,xzb,evan}. 
%\section{Results}

   Because of E864's large acceptance, we are able to divide each measurement 
into several rapidity ($y$) and transverse momentum ($p_{T}$) bins. 
At $y\simeq1.9$ and $p_{T}/A\leq300MeV/c$, the acceptance is about 
20\% for most of the light nuclei. It is in this momentum range where we 
can compare yields of the different species. 
In other bins, we do not have acceptance for 
all the light nuclei detected. Interesting and more detailed 
aspects of the rapidity and transverse momentum distribution of the 
produced light nuclei will be presented in other papers\cite{prc}. 

For charged nuclei with $A\leq3$, the production rates are relatively high 
so that the LET trigger is not required and the analysis does not use 
the hadronic calorimeter. 
%normal trigger condition without LET was used to collect the data and 
%the analysis does not need to include the hadronic calorimeter. 
The calorimeter is the only significant detector for the neutron 
measurement where the tracking detectors serve as tools to calibrate the 
system and to reject the charged particles which deposit energy in the 
calorimeter. For high mass nuclei ($A\geq4$) with low 
production rate, the LET was required to select those events with high mass 
candidates. 
The LET rejects those events without any high mass candidates and achieves a 
rejection factor of about 70 while maintaining good efficiency 
for high mass states 
($\simeq50\%$ for $^{4}He$ and ${}^{>}_{\sim}85\%$ for $A\geq5$).

  Invariant yields are calculated and presented in terms of 
$d^{2}N/(2\pi p_{T}dp_{T}dy)$ in 
units of $GeV^{-2}c^{2}$ as shown in Table~\ref{tab:y_vs_a7}. 
The number of particles (N) are taken from the mass 
distributions in each rapidity and $p_{T}$ bin with background subtracted. 
The background level is about 10\% or less in most of the bins. 
Acceptance and efficiency corrections using either simulation or data 
analysis are applied with the associated systematic and statistical errors. 
The errors of the measurements of proton, neutron, deuteron and $^{3}He$ are 
on the order of 10\%. The rest of the measurements have larger errors 
varying from 20\% ($\alpha$, $^{6}He$) to 45\% ($^{6}Li$, $^{7}Li$, $^{7}Be$) 
due to uncertainty in the LET trigger 
efficiency ($\simeq10\%$), low statistics (up to $40\%$) and high
background level from lower mass nuclei (up to $20\%$). Target 
absorption of the produced particles is negligible for the 5\% Pb target and 
about 8\% for the 60\% Pt target. 
The total systematic error for these measurements varies from a few percent 
to about 25\%. 

\begin{figure}
\centering
\epsfxsize=3.375in \leavevmode 
\epsffile[30 136 534 676]{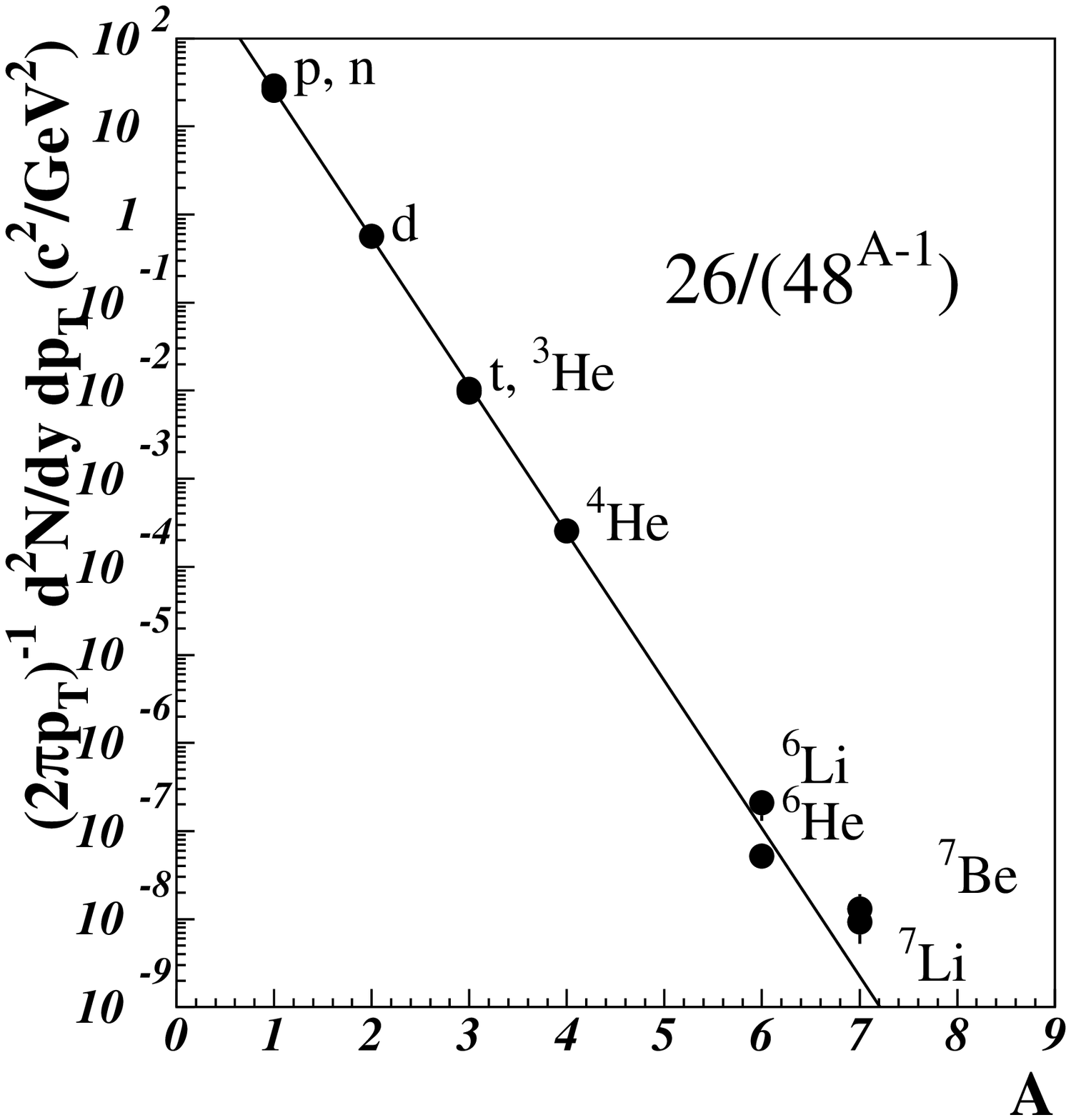}
\caption{{Invariant yield as a function of nuclear number
   A. The bin size in rapidity is 0.2 (0.6 for $^{7}Li$).
   $p_{T}/A\leq 300MeV$. 
   The $J^{P}$'s of p, n, d, $^{3}He$, t, 
   $^{4}He$, $^{6}He$, $^{6}Li$, $^{7}Li$, $^{7}Be$ are ${1 \over 2}^{+}$,
   ${1 \over 2}^{+}$, ${1}^{+}$, ${1 \over 2}^{+}$, ${1 \over 2}^{+}$, 
   ${0}^{+}$,$0^{+}$, ${1}^{+}$, ${3 \over 2}^{-}$, ${3 \over 2}^{-}$
   respectively. The line is a fit to the data with $26/(48^{A-1})$.}}
\label{fig:y_vs_a7}
\end{figure}

  Figure~\ref{fig:y_vs_a7} shows the invariant yield as a function of nuclear 
number A for stable or metastable particle with $A=1$ to $A=7$. 
%(neutron, proton, deuteron, 
%$^{3}He$, triton, $^{4}He$, $^{6}He$, $^{6}Li$, $^{7}Li$ and $^{7}Be$). 
The rapidity binning is $\Delta y=0.2$ with the exception of 
$\Delta y=0.6$ for $^{7}Li$. 
$p_{T}$ binning is $\Delta p_{T}=100MeV$ (100MeV-200MeV) for A=1 to 3, 
$250MeV$ (500MeV-750MeV) for $^{4}He$, $500MeV$ (500MeV-1GeV) for 
$^{6}He$ and $^{6}Li$, and $2GeV$ (0-2GeV, acceptance peaks at 750MeV) for $^{7}Li$ and $^{7}Be$. 
These bin sizes keep $p_{T}/A\leq300MeV$. 
The invariant yields span almost ten orders of magnitudes with striking 
exponential behavior. 

A fit to the A dependence of the invariant yields in this rapidity bin results 
in a penalty factor of about 50 ($48\pm3$, $\chi^{2}/8=4.9$) for each 
additional nucleon to the nuclear cluster. This penalty factor is much 
higher than the penalty factor in the system with lower beam energy at the 
BEVALAC\cite{baltz,bevalac}. 
The consequence is that it is much harder to form high mass objects by 
coalescence, such as strange quark matter.
%which has significant impact on the strangelet search. 

\begin{figure}
\centering
\epsfxsize=3.375in \leavevmode 
\epsffile[38 134 534 626]{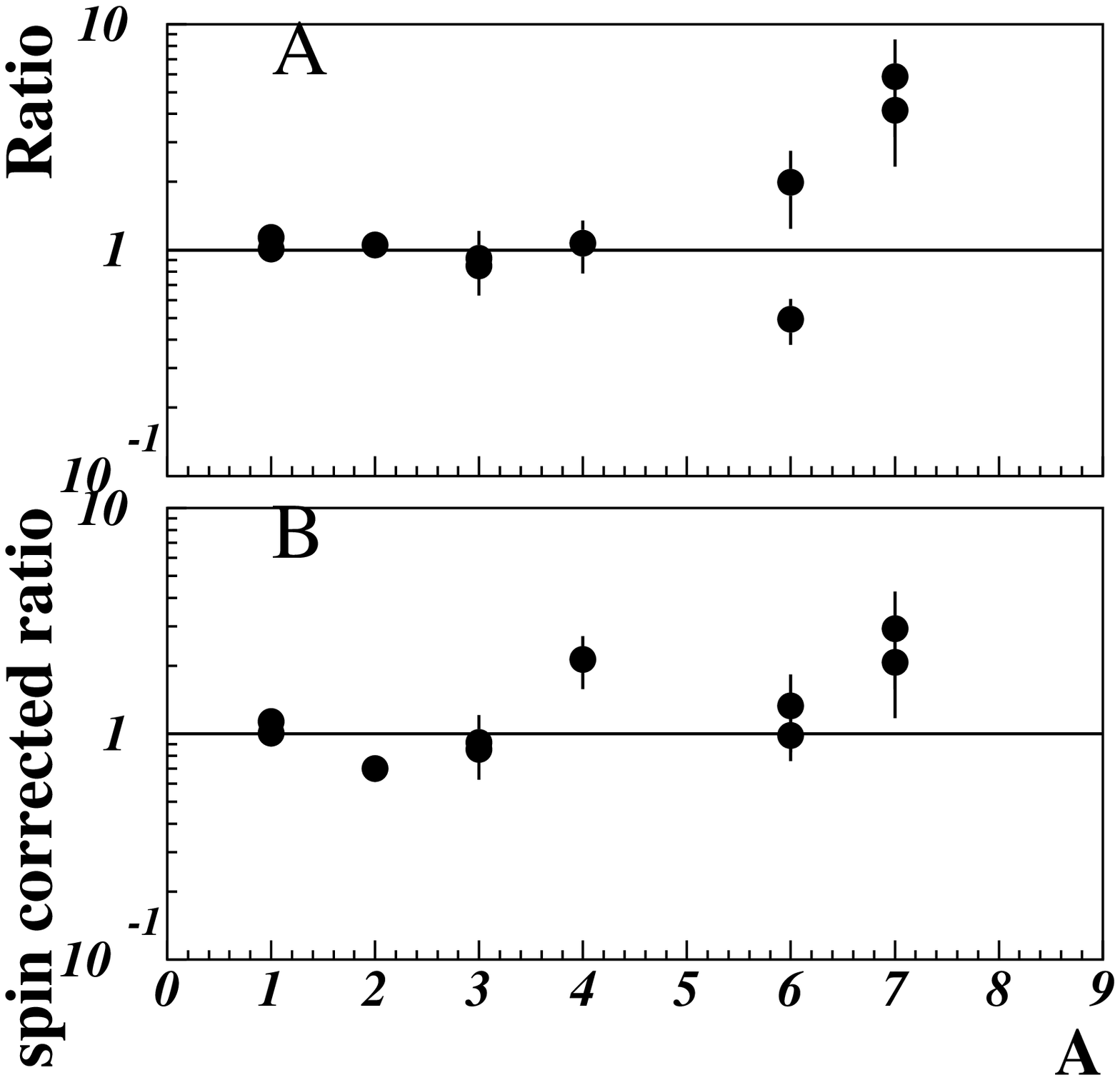}
\caption{%\protect\large
{Ratio of the invariant yield to the exponential function 
vs. nuclear number A with and without spin factor. 
The top panel is the ratio and the bottom panel is the 
ratio with spin correction. See text for details.}}
\label{fig:spin}
\end{figure}

In a statistical approach to the formation of light nuclei, the yield is proportional to the spin factor (2J+1)\cite{nagle,baltz,braun,heinz,mekjian}. 
One can go one step further to analyze the 
deviations of the invariant yields from the exponential behavior 
(figure~\ref{fig:spin}.A). The measured ratios of proton to neutron, 
$^{3}He$ to triton, $^{6}Li$ to $^{6}He$ and their corresponding spin factors 
strongly indicate that the production rate is proportional to the spin 
factor ($2J+1$) of the produced particle\cite{prc,nigel,xzb,evan}.
Therefore, it is probably reasonable to include this spin factor in the 
production rate, which is consistent with most of the 
models\cite{nagle,baltz,braun,heinz,mekjian}. 
Figure~\ref{fig:spin}.B shows the spin 
corrected deviations (from exponential behavior), which still have significant 
deviations. We note, however, that the ratios for $A=6$ states with spin 
factors differing by a factor of 3 are brought into agreement with each 
other by the spin factor correction\cite{prc,xzb}. The spin correction 
factor is taken as $(2J+1)/(2\times{1\over2}+1)$ with a normalization 
to the nucleon spin factor of 2. 
%However, we know that the production rate should depend on the binding energy 
%of the nuclei. Indeed, 

\begin{figure}
\centering
\epsfxsize=3.375in \leavevmode 
\epsffile[30 134 534 672]{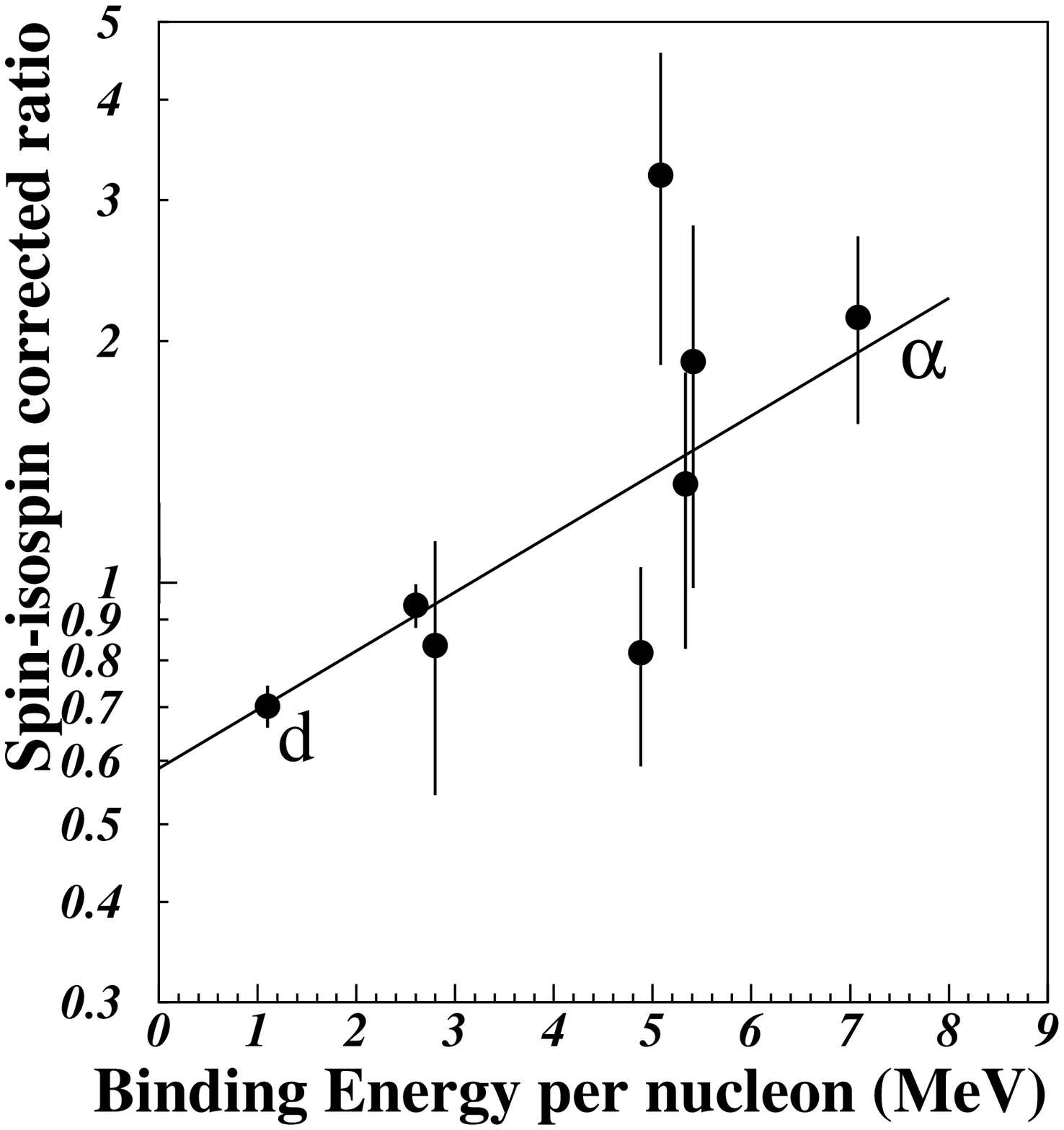}
\caption{%\protect\large
{Ratio of the spin-isospin-factor-corrected invariant yield to the 
exponential function exhibits exponential dependence on the binding energy per 
nucleon. The isospin abundance is (n/p)=1.2. 
Alphas have the largest binding energy per nucleon and therefore have 
the highest production rate deviation from the global exponential behavior. 
Neutrons and protons are not in this plot. Binding energies are calculated 
according to the mass difference between the nucleus and its constituent 
nucleons. See text for details.}}
\label{fig:binding}
\end{figure}

When the deviation after spin factor correction is plotted as a function of 
binding energy per baryon as shown in Figure~\ref{fig:binding}, 
we can fit the dependence with an exponential function of an inverse 
slope of $T_{s}=5.9\pm1.1MeV$ when the A dependence of 
$(2J+1)/2\times26/48^{A-1}$ is applied. 
The small difference (n to p ratio of about $1.2\pm0.1$) 
between the abundances of neutron and proton at 
freeze-out\cite{prc,nigel,evan} is corrected for in the analysis shown 
in Figure~\ref{fig:binding}. If the total binding energy\cite{mekjian} 
instead of the binding energy per baryon is used, the inverse slope is about 
$T_{s}\simeq36MeV$. 
%\section{Conclusion and Discussion}

  From the measurements of light nuclei production near midrapidity 
(at $y\simeq1.9$, where $y_{CM}\simeq1.6$) with low transverse momentum, 
a penalty factor of about 50 for each additional nucleon is found in the 
invariant yield. Although 
the total production rate comes from the integration of the whole phase 
space which might differ from the measured invariant yield (due to flow, 
etc.), we do not expect the ratios of different particle species between 
total production and the rapidity and $p_{T}$ range we cover to be 
different by orders of magnitude. 
In fact, if we use the parametrization of the correction factor from
\cite{heinz}, the penalty factor is estimated to be between 39 and 72. 
Therefore, the penalty factor can be applied to estimate the production 
rates of heavy clusters. Due to the small elastic structure function of 
light nuclei at large momentum transfer\cite{dq1}, the possibility of 
light nuclei at $1.<y<2.2$ coming from projectile or target is small 
compared to the observed production rates.

Naive comparison with theoretical estimates of chemical 
potential and temperature at freeze-out from low mass hadronic 
spectra\cite{braun} will show us whether the light nuclei might have 
different chemical potential and temperature at hadronic freeze-out. 
If we use the chemical potential of  
$\mu_{N}=540MeV$ and temperature of $T=120MeV$\cite{braun}, we get a penalty 
factor of $\exp{[(m_{N}-\mu_{N})/T]}\simeq 28$. 
A kinetic freeze-out and radial flow analysis shows a 
higher penalty factor of about 75 with $\mu_{N}=536MeV$ and 
$T\simeq 93MeV$\cite{heinz}. 

  The dependence of production rate on binding energy per baryon 
shows the sensitivity of our data and can not be 
explained by the coalescence model or the thermal model with the simple 
$\exp{[-B/T]}$ where $T\simeq 100MeV$ and $B$ is the total binding energy
\cite{mekjian}. 
One possible explanation is that the production rate might depend on the 
size of the produced object, and therefore depend on its binding energy. 
It is also possible that there exists some subtle final state interaction 
which depends on the size or the binding energy. For example, collisions 
with sufficient energy to break up the nuclei can occur down to surviving 
temperatures which are comparable with the binding energy\cite{gsi}. 
%On the other hand, the production 
%of pions, kaons, etc. can not occur much below the freeze-out temperature.

  In summary, E864 measures the invariant yields of light nuclei production 
of protons, neutrons, deuterons, $^{3}He$, triton, $^{4}He$, $^{6}Li$, 
$^{6}He$, $^{7}Li$ and $^{7}Be$ near $y_{CM}$ and $p_{T}\simeq0$. A striking 
exponential behavior of light nucleus production as a function of nuclear 
number is found. The penalty factor, extracted from the light nuclei 
production rates, for an additional nucleon in the nuclear cluster is 
about 50. Detailed analysis reveals that the invariant yield may also 
depend on the spin factor and the binding energy of the produced nucleus. 

We gratefully acknowledge the excellent support of the AGS staff. This work
was supported in part by grants from the U.S. Department of Energy's High 
Energy and Nuclear Physics Divisions, and the U.S. National Science Foundation.
\small{
\begin{description}
\item[$\ast$]{Present address: (11)}
\item[$\dag$]{Present Address: Anderson Consulting, Hartford, CT}
\item[$\ddag$]{Present address: Univ. of Denver, Denver CO 80208}
\item[$\S$]{Deceased.}
\item[$\|$]{Present address: McKinsey \& Co., New York, NY 10022}
\item[$\P$]{Present address: Department of Radiation Oncology, Medical College of Virginia, Richmond VA 23298}
\item[$\natural$]{Present address: University of Tennessee, Knoxville TN 37996}
\item[$\sharp$]{Present address: Institut de Physique Nucl\'{e}aire, 91406 Orsay Cedex, France}
\item[$\ell$]{Present Address: Institute for Defense Analysis, Alexandria VA 22311}
\item[$\dag\dag$]{Present Address: MIT Lincoln Laboratory, 
Lexington MA 02420-9185}
\item[$\P\P$]{Present address: (2)}
\end{description}
}

% \begin{figure}
% \caption{}
% \label{}
% \end{figure}

% \begin{table}
% \caption{}
% \label{}
% \end{tabular}
% \end{table}
\pagebreak

\pagebreak
%\onecolumn
%\begin{figure}
%\centering
%\epsfxsize=3.5in \leavevmode 
%\epsfysize=2.5in \leavevmode
%\epsffile[84 336 548 600]{e864.ps}
%\caption{{Schematic views of the E864 spectrometer. In the plan
%view, the downstream vacuum chamber is not shown. M1 and M2 are dipole
%analyzing magnets. S2 and S3 are straw-tube chambers. H1, H2 and H3 are
%scintillator hodoscopes, and CAL is a hadronic calorimeter.}}
%\label{fig:e864}
%\end{figure}

\pagebreak

\pagebreak

\pagebreak

\pagebreak

\pagebreak
\onecolumn
\begin{table}
\caption{%\protect\large
{Invariant yields of light nuclei production at $y=1.9$ and 
$p_{T}/A\leq300MeV/c$. The bin size in y is $\Delta y=0.2$. 
Invariant yields are calculated in terms of $d^{2}N/(2\pi p_{T}dp_{T}dy)$ in 
units of $GeV^{-2}c^{2}$.
See details in text.}}
\label{tab:y_vs_a7}
  \begin{tabular}{|c|c|c|c|c|c|}
  Species& p & n &d & t & $^{3}He$ \\\hline
 Yield& 29.0 & 25.8 & 0.567 & $1.04\times10^{-2}$ & $9.65\times10^{-3}$ \\ 
 ($(GeV/c)^{-2}$)& $\pm3.2$ & $\pm1.7$ & $\pm0.034$ & $\pm6.6\times10^{-4}$ & $\pm3.3\times10^{-3}$ \\
\hline
%\end{tabular}
%  \begin{tabular}{|p{1.5cm}|p{1.5cm}|p{1.5cm}|p{1.5cm}|p{1.5cm}|p{1.5cm}|}
 Species& $^{4}He$ & $^{6}He$ & $^{6}Li$ & $^{7}Li$ & $^{7}Be$  \\
\hline
 Yield& $2.55\times10^{-4}$ & $5.2\times10^{-8}$ & $2.1\times10^{-7}$ &  $0.92\times10^{-8}$ & $1.3\times10^{-8}$ \\
($(GeV/c)^{-2}$)& $\pm6.7\times10^{-5}$ & $\pm1.2\times10^{-8}$ & $\pm7.9\times10^{-8}$ & $\pm4.0\times10^{-9}$ & $(+3.5-2.5)\times10^{-9}$ \\
\end{tabular}
\end{table} 

\end{document}